# OSCILLATION THRESHOLDS FOR "STRIKING OUTWARDS" REEDS COUPLED TO A RESONATOR

**Silva F., Kergomard J., Vergez C.**
*LMA-CNRS, 31 Chemin Joseph Aiguier, 13402 Marseille Cedex 20, France*
*silva@lma.cnrs-mrs.fr*

**Abstract**

This paper considers a "striking outwards" reed coupled to a resonator. This expression, due to Helmholtz, is not discussed here : it corresponds to the most common model of a lip-type valve, when the valve is assumed to be a one degree of freedom oscillator. The presented work is an extension of the works done by Wilson and Beavers (1974), Tarnopolsky (2000). The range of the playing frequencies is investigated. The first results are analytical : when no losses are present in the resonator, it is proven that the ratio between the threshold frequency and the reed resonance frequency is found to be necessarily within the interval between unity and the square root of 3. This is a musical sixth. Actually the interval is largely smaller, and this is in accordance with e.g. the results by Cullen et al.. The smallest blowing pressure is found to be directly related to the quality factor of the reed. Numerical results confirm these statements, and are discussed in comparison with previous ones by Cullen et al (2000).

## INTRODUCTION

In an important paper Wilson and Beavers [1974] studied the oscillation thresholds for a clarinet-type reed exciting a cylindrical duct, using both theory and experiment. They especially showed that the control of oscillations with a playing frequency close to the a duct resonance requires a sufficient damping of the reed. Moreover they showed that the playing frequency is always smaller than the first reed resonance. More recently, studies have been published concerning a more general class of reeds, with a similar theory and new experimental results [Tarnopolsky , 2000 ; Cullen, 2000].

In the present paper we will apply the method of Wilson and Beavers to the case of a "striking outward" reed, and exhibit major differences with a "striking inward" one. This kind of reed corresponds to the most common model of a lip-type valve, when the valve is assumed to be a one degree of freedom oscillator (for a discussion, see e.g. [Cullen, 2000]). Our goal is to derive approximate expressions, exhibiting some basic properties of such reed-tube system.

We first present the model, then the characteristic equation to be solved, finally analytical and numerical results.

## MODELLING A STRIKING OUTWARDS REED

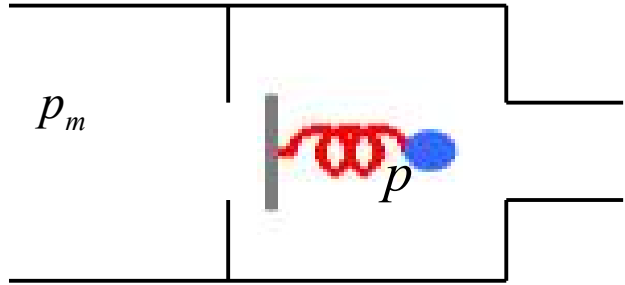

*Figure 1 : schematic representation of a striking outwards reed (after [Fletcher, 1991]*

The model we use is very classical, and extremely simplified. A one degree of freedom oscillator is submitted to pressure forces : a pressure $p_m$ in the mouth, assumed to be constant, and an acoustic pressure $p$ at the input of a resonator. This is described by the following equation:

$$\frac{d^2 y}{dt^2} + q_r \omega_r \frac{dy}{dt} + \omega_r^2 y = \frac{p_m - p}{\mu}. \qquad (1)$$

where $y(t)$ is the reed displacement, $q_r$ the damping parameter, $\omega_r$ the angular eigenfrequency, and $\mu$ the mass per unit length of the reed. For a lip-reed, the displacement has a vertical component, therefore the behavior is much more complicated than this of the spring shown in Fig. 1. Nevertheless we assume the model to be valid. We emphasize that in static regime, if the pressure $p$ is zero, the pressure $p_m$ needs to be negative for the reed to be closed. Following [Wilson, 1974, see also Kergomard, 1995], we chose to use dimensionless variables. The origin for $y(t)$ being the position when $p_m=p=0$, we define $-H$ as the position and $-p_M$ as the pressure for which the reed is closed, H and $p_M$ being positive quantities :

$$p_M = \mu \omega_r^2 H.$$

Writing $\gamma = p_m/p_M$, $x = y/H - \gamma$ and replacing $p/p_M$ by $p$, Eq. (1) becomes :

$$\frac{1}{\omega_r^2}\frac{d^2 x}{dt^2} + \frac{q_r}{\omega_r^2}\frac{dx}{dt} + x = -p \quad . \qquad (2)$$

The following equation is nonlinear : it is the Bernoulli equation, written with certain hypotheses, relating the volume velocity to the opening area of the reed and the pressure difference. We write it directly using dimensionless variables and parameters :

$$u = \zeta(1 + x + \gamma)\sqrt{\gamma - p} \qquad (3)$$

where $\zeta$ is related to both the reed opening and stiffness : $\zeta = Z_c wH \sqrt{\dfrac{2}{\rho p_M}}$. $wH$ is the reed opening at rest, $Z_c = \rho c/S$ the characteristic impedance of the tube. $u$ is the volume velocity multiplied by $Z_c/p_M$. These equations are in accordance with the equations given by Wilson and Beavers, but two signs have to be changed, in order to consider an outwards instead of an inwards reed. Finally, the last equation gives the input admittance of the tube, written in the frequency domain (we use capital letters for variables in the frequency domain) :

$$U(\omega) = Y(\omega)P(\omega) \qquad (4)$$

Equations (2) to (4) give the complete model. In order to get the characteristic equation, we need to rewrite (2) in the frequency domain, and to linearize Eq. (3) :

$$X(\omega) = -D(\omega)P(\omega) \qquad \text{where} \quad D^{-1}(\omega) = 1 + \frac{j\omega q_r}{\omega_r} - \frac{\omega^2}{\omega_r^2}. \qquad (5)$$

and

$$u = \zeta\sqrt{\gamma}(1+\gamma)\left[1 + \frac{x}{1+\gamma} - \frac{p}{2\gamma}\right]. \qquad (6)$$

## CHARACTERISTIC EQUATION

Substracting the d.c. component of (6), and using Eqs. (4) and (5) leads to the characteristic equation :

$$Y(\omega) = \zeta\sqrt{\gamma}\left[-D(\omega) - \frac{1+\gamma}{2\gamma}\right]. \qquad (7)$$

It can be compared to the equation for the case of an inwards reed :

$$Y(\omega) = \zeta\sqrt{\gamma}\left[D(\omega) - \frac{1-\gamma}{2\gamma}\right].$$

The latter equation is in accordance with (Wilson, 1974]). In both cases the volume velocity due to the reed movement is ignored (see [Kergomard, 2000]). The characteristic equation can be solved for a given set of parameters, searching for a complex frequency. This has be done numerically, and the instability threshold of the static regime can be deduced as follows : when the imaginary parts of all solutions but one are positive, all solutions but one are exponentially decreasing, and the instability threshold corresponds to one real frequency, obtained when one parameter varies ζ, γ, $\omega_r$, etc… We do not discuss here the method for solving such equations, and give in the last section some examples of results.

We notice that the instability threshold of the static regime is not always the oscillation threshold : it is true only for direct bifurcation (see e.g. [Grand, 1996]). This question is important, because for an inverse bifurcation, the pressure threshold can be slightly different, but hopefully the playing frequency does not vary too much with the parameters such as the blowing pressure γ. This question remains to be investigated.

## ANALYTICAL APPROXIMATION

Solving Eq. (7) for real frequency gives the instability thresholds, as done by Wilson and Beavers. In order to get some basic results, we first ignore the losses in the resonator, so that $Y(\omega)$ is purely imaginary. Eq. (7) is split into two equations, written for the particular case of a cylindrical tube :

$$\gamma = -\frac{\eta}{2+\eta} \quad (8) \quad ; \quad \cot k_r L \vartheta = -\frac{\zeta\sqrt{\gamma}\vartheta q_r}{\eta(1-\vartheta^2)} \quad (9) \quad \text{where} \quad \eta = 1 - \vartheta^2 + \frac{q_r^2\vartheta^2}{1-\vartheta^2}.$$

$L$ is the length of the resonator, $k_r = \omega_r/c$, and $\vartheta = \omega/\omega_r$. For a given resonator length, the two equations can be solved for the two unkowns, the mouth pressure γ and the playing frequency $\vartheta$, at the threshold. The parameter $\gamma$ can be eliminated from these equations, and the following equation is obtained for $\vartheta$ :

$$\mathrm{Im}(Y) = \zeta q_r G(\vartheta)$$

where $G(\vartheta) = \vartheta \left[ (1-\vartheta^2)^2 + q_r^2 \vartheta^2 \right]^{-1/2} \left[ -(3-\vartheta^2)(1-\vartheta^2) - q_r^2 \vartheta^2 \right]^{-1/2}$     (10).

The last quantity in bracket, in order for the square root to be real, needs to be positive. As a consequence, the playing frequency lies within the following range :

$$\vartheta_1 < \vartheta < \vartheta_2 \quad \text{where} \quad \vartheta_1 \approx 1 + \frac{1}{4}q_r^2 \quad \text{and} \quad \vartheta_2 \approx \sqrt{3}(1 - \frac{1}{4}q_r^2). \quad (11)$$

This important result shows, thanks to this simple model, that for a given reed eigenfrequency, the maximum range for the playing frequency is less than one musical sixth. Is it very different for the case of an inwards reed, for which the range lies between 0 and the reed frequency. There is no "duality" between the two kinds of reed. Moreover, in practice, the interval is shortened because the possibility to get a sound depends on the value of the pressure threshold for a given set of parameters. We notice that the existence of square roots in (10) is related to the Bernoulli equation.

Concerning the pressure threshold, taking now resonator losses into account, it can be shown to depend on both reed damping and tube losses, but the minimum of the threshold when the length of the resonator varies depends mainly on reed damping. This can be seen on the following equation :

$$\gamma_{\min} \approx \gamma_0 \left[ 1 + 2\frac{Y_n}{\zeta}\sqrt{\gamma_0} \right] \quad \text{where} \quad \gamma_0 \approx q_r \left( 1 + \frac{3}{2}q_r \right). \quad (12)$$

$Y_n$ is a minimum (real) value of the input (reduced) admittance of the resonator for a given mode (index $n$). As a consequence, losses in the tube can increase significantly the mimimum of pressure threshold, especially when the parameter $\zeta$ is small. We do not give here neither the derivation of Eq. (12) nor all the analytical results, which will be given in a future article.

## NUMERICAL RESULTS

In what follows, we consider a resonator with a single mode and losses, i.e. with the following expression for the input admittance :

$$Y = Y_n - jQ_n Y_n \frac{1}{j\omega\omega_n}(\omega_n^2 - \omega^2) \ .$$

Solving Eq. (7) by searching for complex frequencies, we are able to get results for various values of the parameters. Figure 2 shows an example of result for the frequency and pressure thresholds, with respect to the ratio $\vartheta_n^{-1} = \omega_r / \omega_n$. When the length or the resonator increases ($\omega_r$ decreases), the threshold frequency decreases to $\vartheta_1$, close to unity (see Eq. (11)), while when the length decreases, the threshold frequency increases to $\vartheta_2$, close to $\sqrt{3}$, and cannot be larger because the threshold pressure tends to infinity. For the chosen values of parameters $q_r$ and $\zeta$, the practical range of possible resonator eigenfrequencies is very narrow, and the minimum value of the threshold pressure lies just above $q_r$ . The influence of the losses in the resonator ($Z_n=1/Y_n$ ) is very small because parameter $\zeta$ is not very small.

Figure 3 shows the same quantities for the case of an striking inwards reed, with the same values of the parameters. For large length of the resonator, the

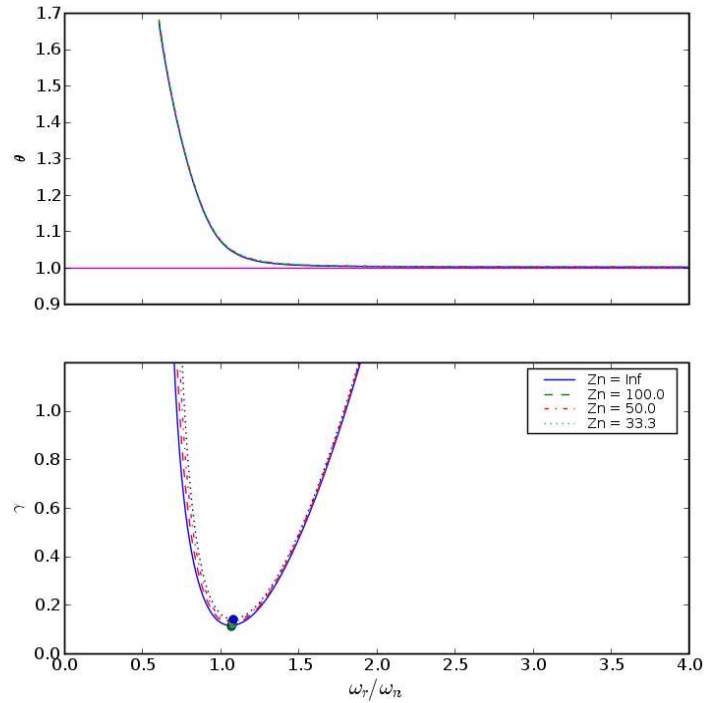

*Figures 2 and 3: results for $f_r$=440Hz ; $q_r$=0.100 ; $\zeta$=0.3, and different values of the impedance peak of the resonator. The abscissa is the ratio of the reed frequency to the resonator frequency. The upper and lower curves represent the ratio of the threshold frequency to the reed frequency, and the threshold pressure, respectively. Fig. 2(above) represents the outwards case, while Fig. 3(below) represents the inwards one.*

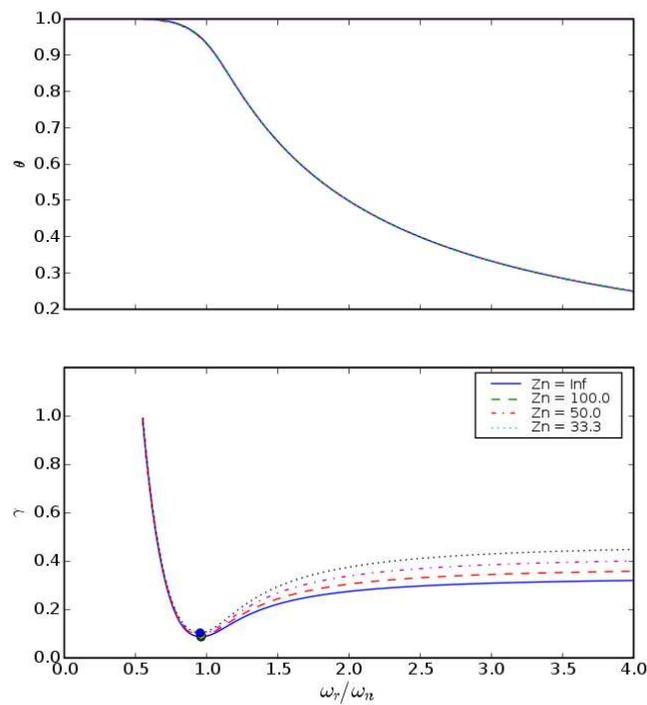

threshold frequency reaches the resonator frequency, the curve becoming hyperbolic, and the pressure threshold tends to a value above 1/3, depending on the losses in the resonator. These properties are well known. For small resonator length, the threshold frequency tends to the reed frequency, but cannot become larger, because the reed becomes beating $(\gamma > 1)$. A similar behavior should be found for the outwards case if negative blowing pressures were considered. Finally the minimum pressure threshold lies just below $q_r$ : one can show that $\gamma_0 = q_r \left( 1 - \frac{3}{2} q_r \right)$. It is smaller than the value for the outwards case. This comparison between the outwards and inwards cases are in qualitative accordance with the work by Cullen et al [2000].

Returning the the outwards case, Figure 4 shows the effect of reed damping, which can be compared to the inwards case (see [Wilson, 1974]) : in both cases weak damping makes easier the oscillation near the minimum value of the threshold pressure, and leads to a threshold frequency closer to the reed frequency.

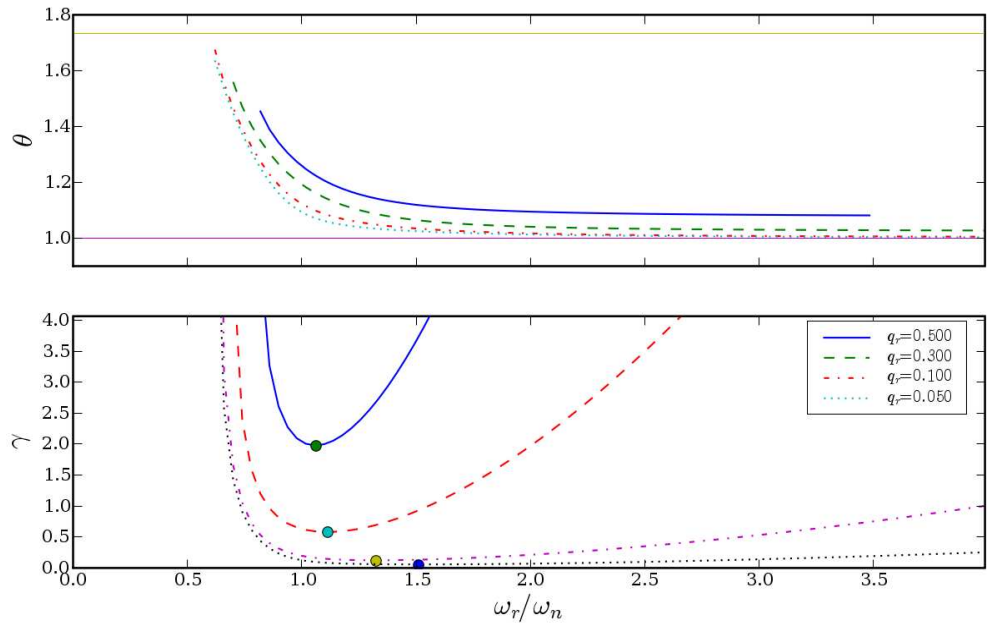

*Figure 4: : results for $f_r$=440Hz ; $\zeta$=0.3 ; $Z_n$=50, and different values of $q_r$.. The abscissa is the ratio of the reed frequency to the resonator frequency. The upper and lower curves represent the ratio of the threshold frequency to the reed frequency, and the threshold pressure, respectively.*

Figures 2 to 4 show results in a similar form than these of Wilson and Beavers. Figure 5 shows results in another form for the case of a cylindrical tube with a fixed length, L, and vaying reed frequency. The effect of the different modes of the resonator can be observed: using the ability to change his lips resonance frequency, the player can keep in tune by bending upward or downward the playing frequency. There are limitations to this action: first, as it can be seen on the upper curve, there exist frequency gaps, i.e. frequencies ranges that cannot be reached without varying the bore length. Then, especially at low frequencies for some values of $\omega_r$, the oscillation threshold is very high, so that a brass player is not able to produce such mouth pressure. Finally, in the higher register, the sensitiveness of playing frequency to lip resonance frequency decreases. Tuning by means of the lips only is more difficult in higher registers than in lower ones, this result corresponding to the feeling of a brass player.

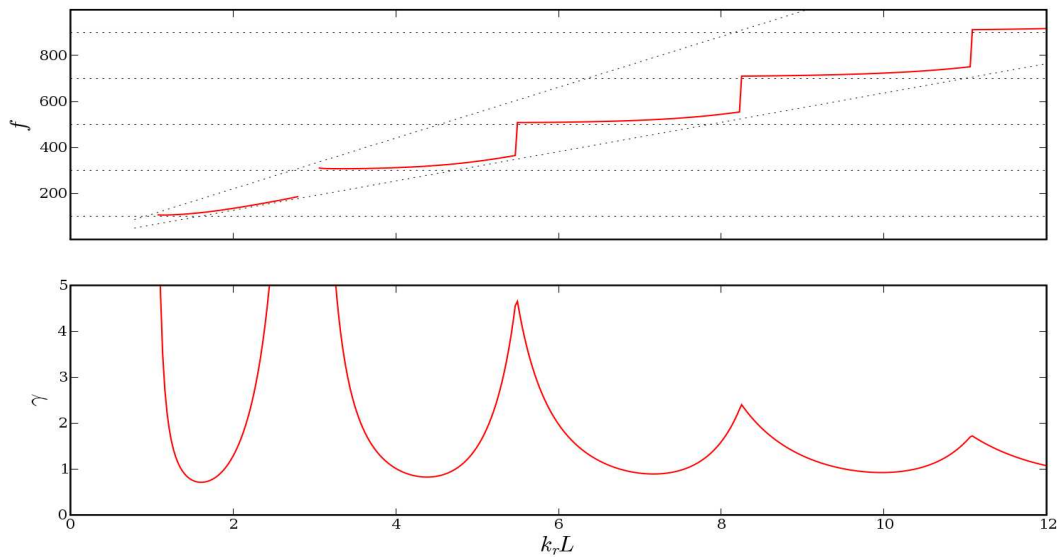

*Figure 5 : results for a fixed length L=1m ; $q_r$=0.3 ; $\zeta$=0.2. The abscissa is $k_rL$, proportional to $\omega_r$. The upper curve gives the threshold frequency, the lower curve gives the threshold pressure.*

## CONCLUSIONS

The present work is based upon previous works, without any novelty concerning the model. What is new is probably the possibility to get analytical results, allowing to understand some essential features, such as the values of the minimum threshold pressures, and the maximum possible value of the threshold frequency compared to the reed eigenfrequency. Further work needs to examine the nature of the bifurcation, using the first harmonic approximation, or better, the small oscillation approach [Grand, 1996].

.